\documentclass[
reprint,
superscriptaddress,
 amsmath,amssymb,
 aps,
]{revtex4-1}
\usepackage{color}

\usepackage{graphicx}
\usepackage{bm}

\begin{document}

\preprint{APS/123-QED}

\title{Anisotropic Ultrafast Spin Dynamics in Epitaxial Cobalt}

\author{Vivek Unikandanunni}
\affiliation{Department of Physics, Stockholm University, SE-10691 Stockholm, Sweden}
\author{Rajasekhar Medapalli}
\affiliation{Center for Memory and Recording Research, University of California San Diego, San Diego, CA 92093, USA}
\affiliation{Department of Physics, School of Sciences, National Institute of Technology, Andhra Pradesh-534102, India}
\author{Eric E. Fullerton}
\affiliation{Center for Memory and Recording Research, University of California San Diego, San Diego, CA 92093, USA}
\author{Karel Carva}
\affiliation{Department of Condensed Matter Physics, Faculty of Mathematics and Physics, Charles University, Ke Karlovu 5,
CZ 121 16 Prague, Czech Republic}
\author{Peter M.\ Oppeneer}
\affiliation{Department of Physics and Astronomy, Uppsala University, P.\,O.\ Box 516, SE-75120 Uppsala, Sweden}
\author{Stefano Bonetti}
\email{stefano.bonetti@fysik.su.se}
\affiliation{Department of Physics, Stockholm University, SE-10691 Stockholm, Sweden}
\affiliation{Department of Molecular Sciences and Nanosystems, Ca' Foscari University of Venice, 30172 Venice, Italy}

\begin{abstract}
We investigate the ultrafast spin dynamics in an epitaxial hcp(1$\bar{1}$00) cobalt thin film. By performing pump-probe magneto-optical measurements with the magnetization along either the easy or hard magnetic axis, we determine the demagnetization and recovery times for the two axes. We observe a $35\%$ slower dynamics along the easy magnetization axis, which we attribute to magneto-crystalline anisotropy of the electron-phonon coupling, supported by our \textit{ab initio} calculations. This points towards an unambiguous and previously undisclosed role of anisotropic electron-lattice coupling in ultrafast magnetism.

\end{abstract}
\maketitle

Ultrafast quenching of magnetic order at sub-picosecond time scales triggered by femtosecond laser pulses and its subsequent recovery were observed in a ferromagnetic nickel thin film in the pioneering experiment by Beaurepaire \textit{et al.}\  \cite{beaurepaire1996ultrafast}. Since then, many experiments have confirmed the occurence of this phenomenon in metallic thin-film ferromagnets \cite{hohlfeld1997nonequilibrium,stamm2007femtosecond,koopmans2005unifying,dalla2007influence,
carpene2008dynamics,malinowski2008control,boeglin2010distinguishing,koopmans2010explaining,chan2012ultrafast,kirilyuk2010ultrafast,radu2011transient,ostler2012ultrafast,mathias2012probing,rudolf2012ultrafast,turgut2013controlling,mangin2014engineered,lambert2014all,Turgut2016}. Significant theoretical progress \cite{zhang2000laser,djordjevic2007connecting,steiauf2009elliott,krauss2009ultrafast,battiato2010superdiffusive,chimata2012microscopic,carva2013ab} has been made towards finding the fundamental microscopic mechanisms able to explain how angular momentum is lost and recovered at these ultrafast time scales, orders of magnitude faster than expected, e.g. by the textbook Landau-Lifshitz theory \cite{landau1965collected}. Despite a two-decade long quest, a complete understanding of the phenomenon is still lacking.

The role of the lattice in ultrafast magnetism has been discussed since the early years following the pioneering experiment. The Elliott-Yafet-type spin-flip scattering was put forward as a possible mechanism through which angular momentum can be transferred from the spin system to the lattice, although the efficiency of this mechanism has been debated \cite{krauss2009ultrafast,steiauf2009elliott,bigot2009coherent,steiauf2010extension,carva2011ab}. Surprisingly, only very few experimental studies \cite{koopmans2000ultrafast,vomir2005real,bigot2005ultrafast,henighan2016generation,dornes2019ultrafast} have investigated epitaxial systems, where the crystalline structure of the sample can be properly modeled. Recently, using femtosecond x-ray diffraction, it was observed that a femtosecond optical pulse can trigger ultrafast coherent terahertz longitudinal acoustic phonons (up to 4 THz) in an epitaxial iron thin film \cite{henighan2016generation}, disproving the common assumption that the lattice cannot respond coherently on ultrafast time scales. Even more recently, another ultrafast x-ray experiment on a similar iron film \cite{dornes2019ultrafast} suggested the possibility of the ultrafast version of the Einstein-de Haas experiment, where the demagnetization of the material is compensated by a coherent mechanical rotation of the body, in this case, driven by generation of transverse acoustic phonons at terahertz frequencies. Notwithstanding, the unambiguous detection of the involvement of the lattice structure in ultrafast magnetization dynamics is still to be achieved.

In this Letter, we investigate a different model system, 
an epitaxial hcp(1$\bar{1}$00) cobalt thin film, which we probe with a femtosecond optical pump-probe setup. Using the time-resolved magneto-optical Kerr effect (TR-MOKE), we measure the demagnetization and recovery of the sample magnetization on the femto- and picosecond time scales. The magnetization is set along the easy or hard magnetization axis, which corresponds to two orthonormal lattice directions in the thin film plane. 
Surprisingly, there do not yet exist systematic studies on possible anisotropic ultrafast spin dynamics in epitaxial model systems with strong magnetocrystalline anisotropy, which would allow to pin-down the role of the orientation-dependent electron-phonon coupling. In the following, we show that our measurements are able to reveal distinct magnetization dynamics coupled to the anisotropic lattice structure, even without the atomic resolution given by an x-ray probe, and that is consistently explained by \textit{ab initio} calculations of the anisotropic electron-phonon interaction.


\begin{figure}[hbt!]
\includegraphics[width=\columnwidth]{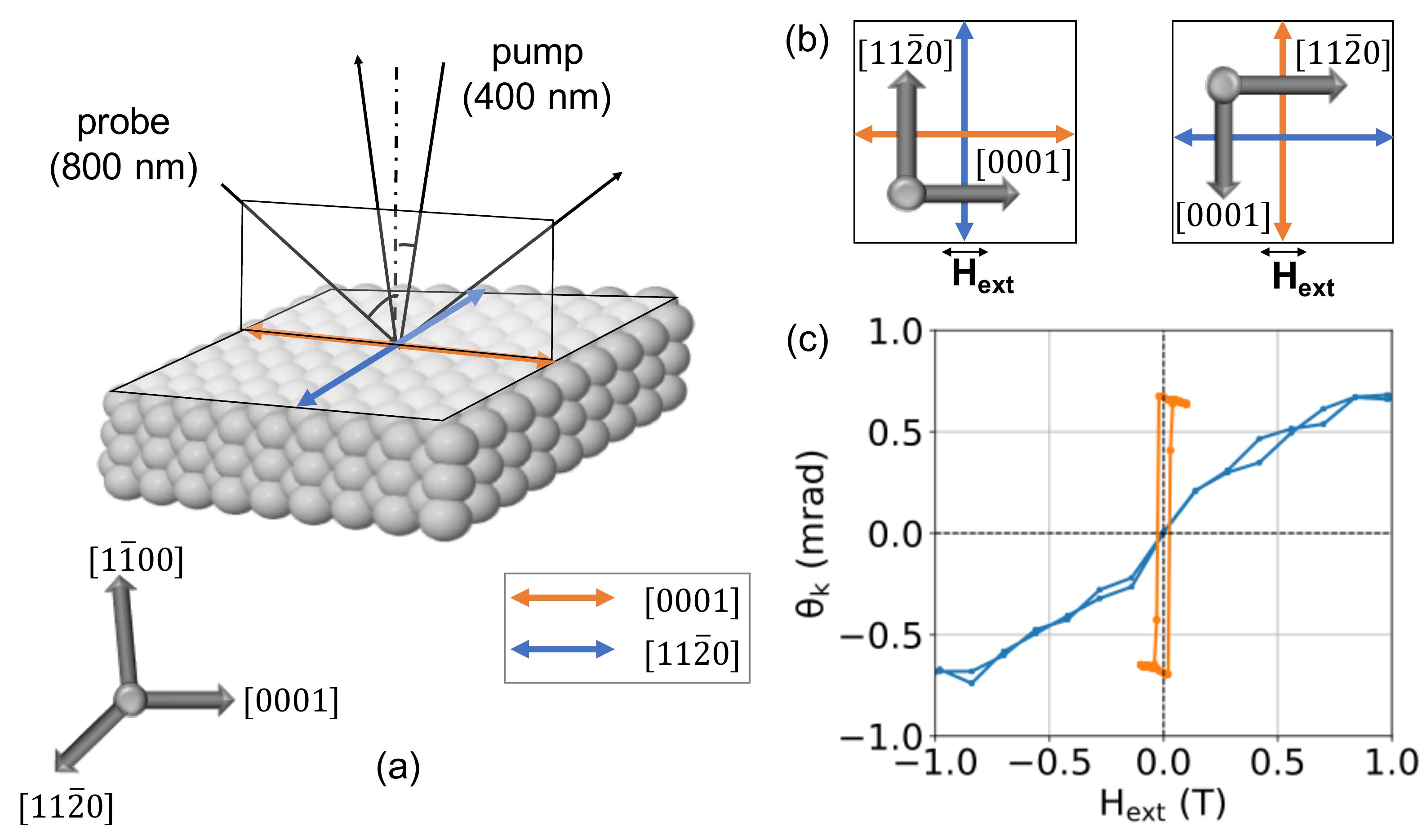}
\caption{\label{fig:fig1}(a) Crystallographic directions of the hcp Co film and geometry of the MOKE setup. (b) Relative orientation of the sample with respect to the externally applied magnetic field \textbf{H}$_{\rm ext}$ parallel to (left panel) the [0001] easy magnetization axis and (right panel) to the [11$\bar{2}$0] hard magnetization axis. (c) Magnetization loops along the easy (orange) and hard (blue) axes measured using the longitudinal MOKE.}
\end{figure}

A 15-nm-thick epitaxial hcp(1$\bar{1}$00)-cobalt thin film was grown as Co[1$\bar{1}$00] on a MgO(110) substrate and a Cr(211) seed layer. The Co layer was capped with a 3-nm-thick Pt layer. The easy axis of magnetization is along the $c-$axis [0001] and lies in the plane of the film, as shown in Fig.~\ref{fig:fig1}(a). The hard axis of magnetization [11$\bar{2}$0] is perpendicular to it and also in the sample plane. This strong in-plane magnetic anisotropy of the film enabled us to measure the ultrafast demagnetization along two different crystalline orientations by a simple in-plane rotation of the sample, as shown schematically in Fig.~\ref{fig:fig1}(b). Magneto-optical loops along the easy (orange) and hard (blue) magnetization axes measured 
with the longitudinal MOKE are shown in Fig.~\ref{fig:fig1}(c). The loops are qualitatively similar to the vibrating sample magnetometer data presented in the Supplemental Material (SM) \cite{SM}. The probing configuration and laser setup are the same as for the time-resolved data shown below. In this way, the time-resolved data can be directly normalized with respect to the magnetization loops. 

The pump-probe experiments were performed with an amplified Ti:Sapphire laser, with a temporal resolution of approximately 40 fs, repetition rate of 1 kHz and a central wavelength of 800 nm. As pump, we used the 400-nm optical pulses generated by frequency doubling the fundamental of the laser, using a $\beta$-barium borate crystal. As probe, we used the fundamental of the laser at 800 nm. The pump was incident at an angle $\theta_{\rm pump}\approx 10$ degrees and the probe at an angle $\theta_2\approx 55$ degrees, as shown in Fig.~\ref{fig:fig1}(a). In this configuration, we are optimized to measure the longitudinal MOKE, proportional to the in-plane component of the magnetization \cite{zvezdin1997modern}. In addition, the different pump and probe energies allow us to suppress coherent optical artefacts due to the formation of transient gratings in the film \cite{luo2009eliminate}.

\begin{figure}[t]
\includegraphics[width=\columnwidth]{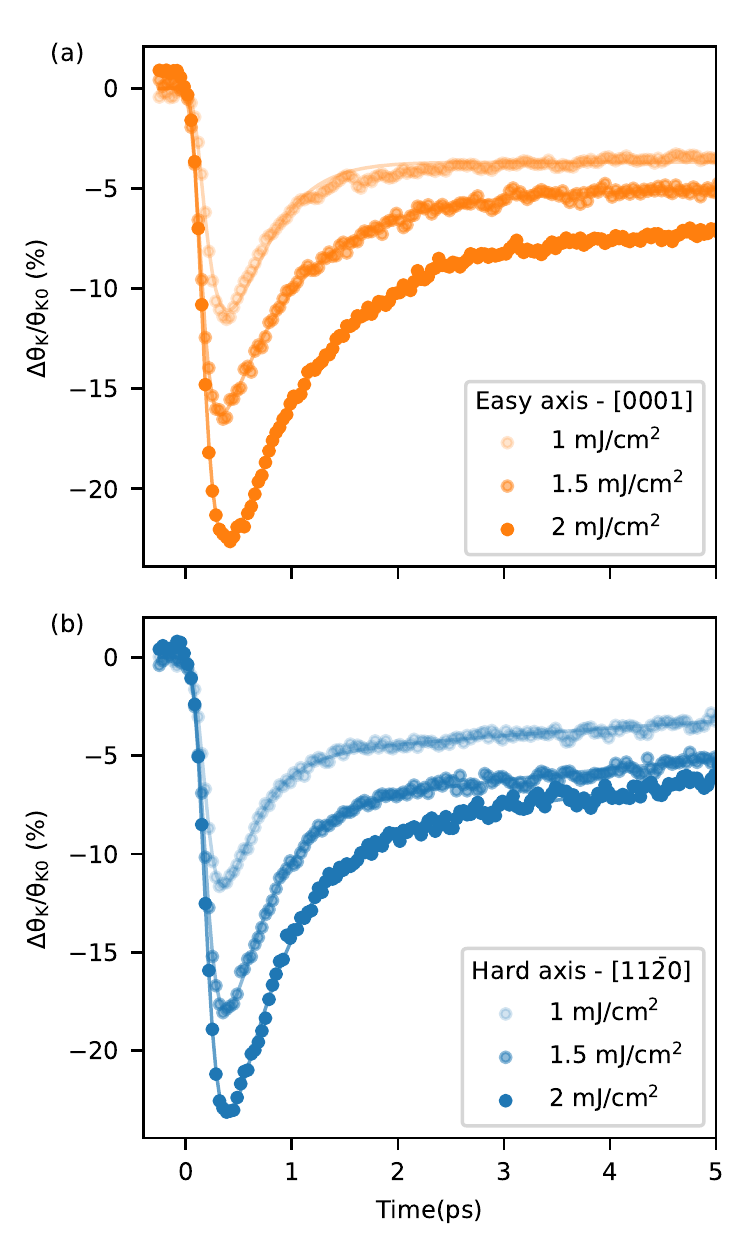}
\caption{\label{fig:fig2}Transient magnetization dynamics in  the hcp Co sample measured along the (a) easy and (b) hard magnetization axes. The pump is \textit{s}-polarized and the probe is \textit{p}-polarized. The calculated fluence absorbed by the film is given in the legend.}
\end{figure}

Figures~\ref{fig:fig2}(a)-(b) show the TR-MOKE measurements performed at selected fluences for the easy and hard magnetization axes. The delay traces are calculated as the difference of the delay traces recorded using magnetic fields of equal magnitude but opposite sign, and then normalized by the amplitude of the magnetization loops shown in Fig.~\ref{fig:fig1}(c). The applied magnetic fields were $\pm400$ mT and $\pm1000$ mT for the easy and, respectively, hard magnetization axis, enough to reach saturation. These values were chosen to result in the same effective field, as demonstrated by the measurement of the sample FMR, see SM \cite{SM}.
We have however checked that none of the observations reported below depend on the magnitude of the external field, given that the sample is saturated. Plotting the difference of opposite fields allows for isolating the pure transient magnetic signal, and remove the contribution from transient reflectivity signal. The transient reflectivity is shown in the SM \cite{SM} and has a maximum relative change of 0.2\% for the highest fluence. This value has to be compared with the maximum relative change in Kerr rotation of about 20\%. The much larger variation indicates that the Kerr signal is measuring genuine magnetization dynamics and not optical artifacts, i.e.\ $\Delta \theta_K(t)/\theta_{K0}=\Delta M(t)/M_0$ \cite{prasankumar2016optical}. Furthermore, we checked the dynamical Kerr rotation and ellipticity of all combinations of \textit{s} and \textit{p} pump and probe polarizations \cite{koopmans2000ultrafast,guidoni2002magneto,carpene2013measurement}, and the MOKE response stayed the same in shape and amplitude, within the experimental uncertainty.

The demagnetization curves shown in Fig.~\ref{fig:fig2} illustrate the typical response observed in this experiment: a rapid quench of the magnetization on a time scale of the order of $\sim100$ fs followed by a fast recovery on the time scale of $\sim1$ ps, and finally a much slower recovery on the 10 to 100 ps time scales. The figure also shows that the maximum demagnetization increases monotonically with the absorbed fluence.
\begin{figure}[t]
\includegraphics[width=\columnwidth]{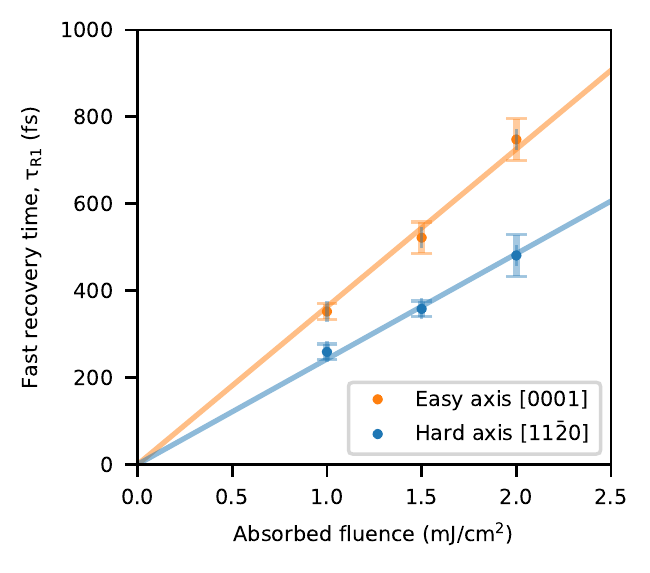}
\caption{\label{fig:fig3} 
Fluence dependence of the fast recovery time $\tau_{R1}$ of the magnetization for (orange) easy and (blue symbols) hard magnetization axes. Solid lines are linear fits to the data, imposing the crossing of the origin of the plot.} 
\end{figure}
In order to accurately determine the demagnetization and recovery time constants, we fitted the ultrafast demagnetization data with the equation given in Ref.~\cite{dalla2008laser}. In 
those equations, the dynamics is described by the decay time $\tau_m$ for the ultrafast demagnetization, $\tau_{R1}$ for the fast recovery, and $\tau_{R2}$ for the slow recovery, 
which we can extract by a careful fitting procedure described in the SM \cite{SM}.

For the three fluences presented in Fig.~\ref{fig:fig2}, there is only negligible difference (i.e.\ within the error bars) in the demagnetization amplitude at each fluence for the two different magnetization orientations. The change of demagnetization time constant $\tau_m $ for these fluences is below the resolution of our measurement \cite{koopmans2010explaining}, and we obtained the best fit with $\tau_m = 130$ fs for all these measurements. By using this value for $\tau_m$, we could reliably extract the fast recovery time constant $\tau_{R1}$ and the slow recovery time constant $\tau_{R2}$ for all the measurements. We also note that the slow recovery time $\tau_{R2}$ is a coarse approximation of the dynamics, and which excludes the full response of the magnetization including the ferromagnetic resonance. Hence, we do not discuss it further in the following.

\begin{figure}[t]
\includegraphics[width=\columnwidth]{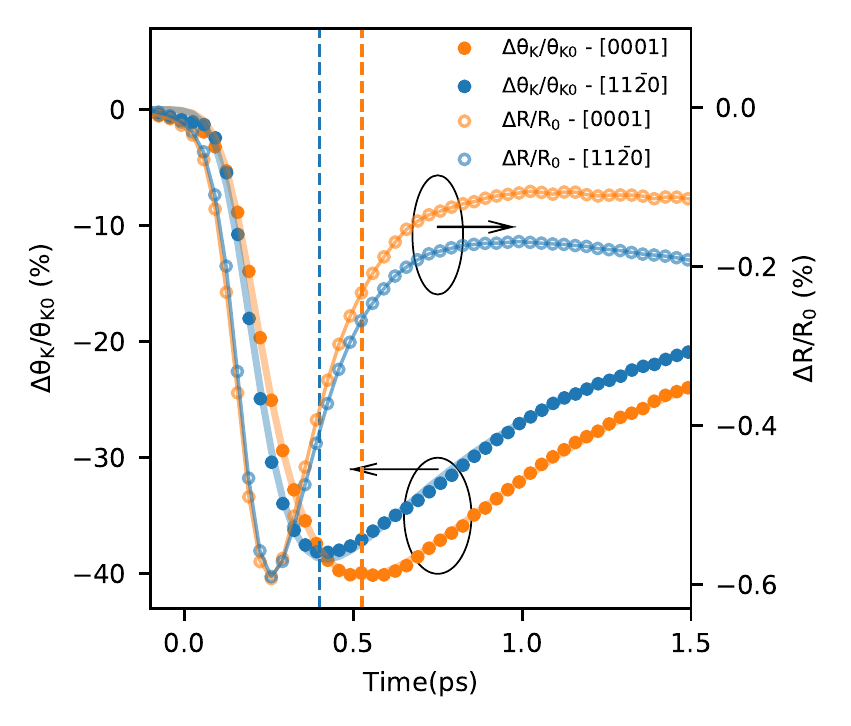}
\caption{\label{fig:fig4}Transient reflectivity (open symbols) and ultrafast magnetization dynamics (solid symbols) measured with the externally applied magnetic field parallel to the easy or hard axis of magnetization in epitaxial hcp cobalt, for an absorbed fluence of approximately 4 mJ/cm$^2$. The dotted lines show the 
time point of maximum demagnetization.}
\end{figure}
We instead focus on the fast recovery time $\tau_{R1}$. Fig.\ \ref{fig:fig3} shows the extracted $\tau_{R1}$ for both orientations as a function of the absorbed fluence. The fast recovery time increases with increasing fluence, consistent with previous reports \cite{koopmans2010explaining,atxitia2010evidence,von2015nonlocal}. In addition, we also observe that the fast recovery of the magnetization along the hard axis orientation is always faster than for the easy axis orientation. The data can be fitted assuming a linear dependence and forcing the fit to go through the origin. The slopes of the lines are approximately 360 fs cm$^2$/mJ for the easy magnetization axis, and 240 fs cm$^2$/mJ for the hard magnetization axis. Hence, and this is one of the main experimental findings of this work, the ultrafast magnetization dynamics in hcp Co recovers systematically \textit{faster} along the hard magnetization axis than along the easy axis. In the absorbed fluence range up to 2 mJ/cm$^2$, the ratio of the easy/hard slopes is approximately 1.5, \textit{i.e.}\ the recovery is approximately 50\% slower along the easy magnetization axis. One could argue that, if this is due to an intrinsic non-equilibrium spin-scattering mechanism within the material, one would expect not only the magnetization recovery to be faster, but also the quenching. However, the expected change of the demagnetization time constant at these fluences is comparable with the resolution of our measurement, and such difference may be not measurable within our experimental resolution.

In order test this hypothesis, we looked at slower demagnetization by increasing the maximum absorbed fluence by a factor two, i.e.\ 4 mJ/cm$^2$, close to the sample damage threshold, observed as a permanent change in sample reflectivity and magneto-optical signal. The data recorded at such fluence is shown in Fig.~\ref{fig:fig4} for the two magnetization axes, where we also show the corresponding transient reflectivity. With the demagnetization process slowed down, we can now resolve the different time constants of the quenching, with $\tau_{m,hard}\approx130$ fs and $\tau_{m,easy}\approx180$ fs, showing that indeed the hard magnetization axis has an overall faster dynamics than the easy axis. We also note that in this measurement $\tau_{m,easy}/\tau_{m,hard}\approx\tau_{R1,easy}/\tau_{R1,hard}\approx1.38$, suggesting that the same microscopic mechanism is governing the quenching and relaxation processes. This ratio is slightly smaller than in the lower fluence cases, but it confirms that the dynamics along the easy magnetization axis is still substantially slower than the one along the hard axis. This can be due to a saturation regime approaching, or simply to sample non-uniformity. In order to check the robustness of our results, we performed several measurements weeks apart, with the sample and the setup both dismounted and mounted again, and the absorbed fluence varied between 1 and 2 mJ/cm$^2$, far away from the damage threshold. Over 10 independent measurements for each axis, we found that the magnetization dynamics along the easy axis recovers always slower than the hard axis one, by $35\pm19\%$. A summary of this data is shown in the SM \cite{SM}.
 
We now turn to the transient reflectivity data shown in Fig.~\ref{fig:fig4}. Along both the [0001] (easy) and [11$\bar{2}$0] (hard) axes, there is a measurable delay between the maximum change in reflectivity and the maximum change in magnetization, by approximately 250 fs for the easy magnetization axis, and 150 fs for the hard magnetization axis. This is similar to what was observed in Ref.~\cite{carpene2008dynamics} for bcc iron pumped with 800 nm pulses and probed at 
shorter wavelengths ($500-540$ nm). As stated in that work, this indicates that the spin dynamics follows the onset of a non-equilibrium electronic distribution.

In order to explain the magneto-crystalline anisotropy of the de- and re-magnetization time constants, we consider first the overall picture for energy transfer from the laser-excited conduction electrons. The conduction electrons thermalize within about 100 fs and transfer their energy to the cold phonons due to electron-phonon coupling. On a time scale longer than the initial electron thermalization, this energy transfer process is reasonably well described by the two-temperature model \cite{Anisimov1974}. The rate of increase of the lattice temperature is given by
$C_{ph} \, \partial T_{ph}/\partial t = G (T_e - T_{ph})$ where $T_e$ ($T_{ph}$) is the electron (lattice) temperature, $C_{ph}$ is the lattice heat capacity, and $G$ is the electron-phonon coupling constant \cite{Grimvall1981,Allen1987}.  This quantity can be computed \textit{ab initio}; it is given by \cite{Grimvall1981}
\begin{equation}
 G = 2\pi g(\epsilon_F)\hbar k_{B}
 \int_0^{\infty} d\Omega \, \alpha^2 F(\Omega) \,\Omega \, ,
 \end{equation}
where $g (\epsilon_F)$ is the electronic density of states at the Fermi level $\epsilon_F$, 
and $\alpha^2F(\Omega)$ is the Eliashberg function, with $\Omega$ the phonon frequency variable (for explicit expressions, see \cite{carva2011ab,carva2013ab}). 

To investigate the dependence of $G$ on the magnetization axis, we have calculated the phonon spectra and electron-phonon matrix elements of hcp Co self-consistently,  using the ELK full-potential code \cite{elk}. Note that the spin-orbit interaction was included, which allows us to examine the influence of the magnetization axis on $G$. We find a significant magneto-crystalline dependence of $G$: for magnetization along [0001] the calculated coupling was $G=1.8 \times 10^{18}$ Wm$
^{-3}$K$^{-1}$, while for magnetization along [11$\bar{2}$0]  it was $2.8 \times10^{18}$ Wm$^{-3}$K$^{-1}$. Hence, the electron-phonon coupling for $M$ along the hard axis is about 50 \% larger as for $M$ along the easy axis.  

Analyzing next where the magneto-crystalline anisotropy in $G$ comes from, we found that the differences between the phonon dynamical matrices and phonon spectra computed for the hard and easy magetization axes are small (see the SM \cite{SM} for the computed phonon spectra). However, we found that the electron-phonon interaction is much more sensitive to the magnetization direction, due to small shifts of the energy levels near the Fermi energy induced by spin-orbit coupling. The biggest changes in the electron-phonon interaction were obtained for high energy phonons.

The implication of the larger electron-phonon coupling $G$ for the hard magnetization axis is a stronger transfer of energy from hot electrons to cool phonons, and thus a faster electron cooling, leading to a faster remagnetization, which is  fully consistent with the  magnetization dynamics measured for $t> 0.5$ ps.  
The high magneto-crystalline anisotropy of the remagnetization  rate corroborates that the recovery is driven by a mechanism that depends strongly on spin-orbit interaction, as the Elliott-Yafet electron-phonon spin-flip scattering. 

An accurate description of the demagnetization dynamics in the first few hundred fs is however a more complex issue. Note that it is  observed only at very high fluences.
Considering an electron-phonon picture, the transfer of spin angular momentum from the electrons to the phonons is given by the Elliott-Yafet electron-phonon spin-flip scattering \cite{koopmans2010explaining}, which is exactly described by the spin-flip Eliashberg function (see \cite{carva2011ab,carva2013ab}). This quantity has a very similar spectral dependence as the conventional
$\alpha
^2 F(\Omega)$ but it is about 40 times smaller \cite{carva2013ab}. It has nevertheless the same magneto-crystalline anisotropy as the  common
$\alpha^2F$. The electron-phonon spin-flip scattering for $M$ along the hard axis is thus larger, which would imply a faster magnetization decay in the first few hundred fs, consistent with our measurements. It needs to be emphasized, though, that in this time interval there will be nonthermal electron populations that depend on the used fluence and several nonequilibrium processes that can be involved, which would limit the validity of the two-temperature model as well as of the here-used quasi-equilibrium electron-phonon scattering description. We can therefore only conclude that the right trend is given on the very short time scale. 

In conclusion, we performed ultrafast magneto-optical pump-probe experiments on epitaxial hcp cobalt, in order to measure the magnetization dynamics along the easy and hard magnetization axes. We observed a systematic 35\% slower quenching and relaxation dynamics along the easy magnetization axis. Our \textit{ab initio} calculations reveal a large magneto-crystalline anisotropy in the electron-lattice coupling and the Elliott-Yafet spin-flip scattering, which explains the observed anisotropic magnetization dynamics.
Our study furthermore introduces a novel approach to probe, using wavelengths in the optical range, the role of the lattice anisotropy in ultrafast magnetism. We envision that future experiments that mimic our approach will be able to explore other crystalline materials with well-defined lattice structures.
The investigation of model systems, as opposed to polycrystalline ones, allows moreover for theoretical models to be tested to a greater accuracy. We anticipate that such studies may give important hints towards completely solving the question of the dissipation of angular momentum at ultrafast time scales, which is yet not settled after more than two decades of research.

We gratefully acknowledge B. Wehinger for useful discussion. V.U.\ and S.B.\ acknowledge support from the European Research Council, Starting Grant 715452 “MAGNETIC-SPEED-LIMIT”. R.M.\ and E.E.F.\ were supported by U.S.\ Department of Energy, Office of Science, Office of Basic Energy Sciences, under Contracts No.\ DE-SC0018237. P.M.O.\ acknowledges support from the Swedish Research Council (VR), the K.\ and A.\ Wallenberg Foundation (Grant No.\ 2015.0060), the CRC/TRR 227 
``Ultrafast Spin Dynamics”, and the Swedish National Infrastructure for Computing (SNIC). K.C.\ acknowledges support from the Czech Science Foundation (Grant No.\ 18-07172S) and The Ministry of Education, Youth and Sports Large Infrastructures for Research, Experimental Development and Innovations project 
``e-Infrastructure CZ – LM2018140“.

\nocite{cullity2011introduction}
\bibliography{Co_anisotropy.bib}

\end{document}


\preprint{APS/123-QED}

\title{Supplemental Material\\Anisotropic Ultrafast Spin Dynamics in Epitaxial Cobalt}

\author{Vivek Unikandanunni}
\affiliation{Department of Physics, Stockholm University, SE-10691 Stockholm, Sweden}
\author{Rajasekhar Medapalli}
\affiliation{Center for Memory and Recording Research, University of California San Diego, San Diego, CA 92093, USA}
\affiliation{Department of Physics, School of Sciences, National Institute of Technology, Andhra Pradesh-534102, India}
\author{Eric E. Fullerton}
\affiliation{Center for Memory and Recording Research, University of California San Diego, San Diego, CA 92093, USA}
\author{Karel Carva}
\affiliation{Department of Condensed Matter Physics, Faculty of Mathematics and Physics, Charles University, Ke Karlovu 5,
CZ 121 16 Prague, Czech Republic}
\author{Peter M.\ Oppeneer}
\affiliation{Department of Physics and Astronomy, Uppsala University, P.\,O.\ Box 516, SE-75120 Uppsala, Sweden}
\author{Stefano Bonetti}
\email{stefano.bonetti@fysik.su.se}
\affiliation{Department of Physics, Stockholm University, SE-10691 Stockholm, Sweden}
\affiliation{Department of Molecular Sciences and Nanosystems, Ca' Foscari University of Venice, 30172 Venice, Italy}

\maketitle

\onecolumngrid

\section{Fitting procedure}

In this section, we describe the detailed fitting procedure, and report all the extracted values. We stress again that a careful fitting with properly chosen conditions is key to obtain meaningful parameters in cases like this, where the number of free parameters is larger than those needed to fit the data. 
In order to accurately determine the demagnetization and recovery time constants, we fitted the ultrafast demagnetization data with the equation given in Ref.\ \cite{dalla2008laser}, namely
\begin{eqnarray}
\frac{\Delta M(t)}{M_0} &=& \Bigg(\frac{A_1\tau_{R1}-A_2 \tau_{m}}{\tau_{R1}-\tau_{m}}e^{-t/\tau_{m}} - \frac{\tau_{R1}(A_1-A_2)}{\tau_{R1}-\tau_{m}}e^{-t/\tau_{R1}} - \frac{A_2}{\sqrt{t/\tau_{R2}+1}}\Bigg)\circledast G(t),
\label{eq:one}
\end{eqnarray}
where each parameter is defined as follows. $\Delta M(t)/M_0$ is the relative change in magnetization. Here, $\Delta M(t)$ is the pump-induced change in magnetization measured with the time-resolved magneto-optical Kerr effect (MOKE), while $M_0$ is proportional to the saturation magnetization, measured as the maximum Kerr rotation when the external magnetic field applied to the sample is large enough to saturate the sample. $\tau_m$ is the demagnetization time constant, $\tau_{R1}$ and $\tau_{R2}$ are the fast and, respectively, the slow recovery time constants. $A_1$ and $A_2$ are adimensional constants related to the magnitude of the demagnetization. The entire expression within the round brackets is convoluted with a Gaussian-shaped temporal profile $G(t)$ which accounts for the finite duration of the  probe pulse.

There are totally five free fitting parameters (A$_1$, A$_2$, $\tau_m$, $\tau_{R1}$ and $\tau_{R2}$) in Eq.~(\ref{eq:one}), since the width of the Gaussian function is fixed and set equal to the experimentally measured optical autocorrelation of the laser pulse, in our case 40 fs. We stress that a fit to the data performed using this equation with all the parameters running free, can produce reasonable $\chi^2$ values but at the same time return recovery time constants that are clearly at odds with the experimental evidence. This problem can be avoided by reducing the number of free parameters, considering the constraints imposed by the realistic physical conditions. We first noticed that $\tau_m$ is extracted reliably even when all parameters are free running, and we noticed that a value $\tau_m = 130$ fs was able to accurately fit all the measurements done in the low to medium fluence regime ($\le$ 2 mJ/cm$^2$). This fact is consistent with the observation of Ref. \cite{koopmans2010explaining}, where they estimated, in this fluence range, a maximum change of $\tau_{m}$ of less than 40 fs, which is also our experimental resolution. For this fluence range, we could hence fix $\tau_{m}=130$ fs, and bring the number of free parameters to four.

We allowed again the fit to run free with now four parameters and looked at the adimensional amplitudes A$_1$ and A$_2$. The parameters A$_1$ is related to the maximum demagnetization amplitude, but not the demagnetization amplitude itself, whereas the parameter A$_2$ is the demagnetization amplitude after the fast recovery. With the four free parameters, we extracted the values of A$_1$ and A$_2$, and we checked that the extracted value of A$_2$ corresponded to the observed demagnetization amplitude after the fast recovery. After this step, we allowed A$_1$ and A$_2$ to vary only within the error of the measurement, substantially fixing them.

Finally, with only two free parameters left, namely $\tau_{R1}$ and $\tau_{R2}$, we could get a robust and reliable fit of the experimental data using Eq. \eqref{eq:one} consistent with the experimental observations. All the parameters extracted are reported in Table \ref{tab:table1}.

\begin{figure*}[!hbt]
\includegraphics[width=\columnwidth]{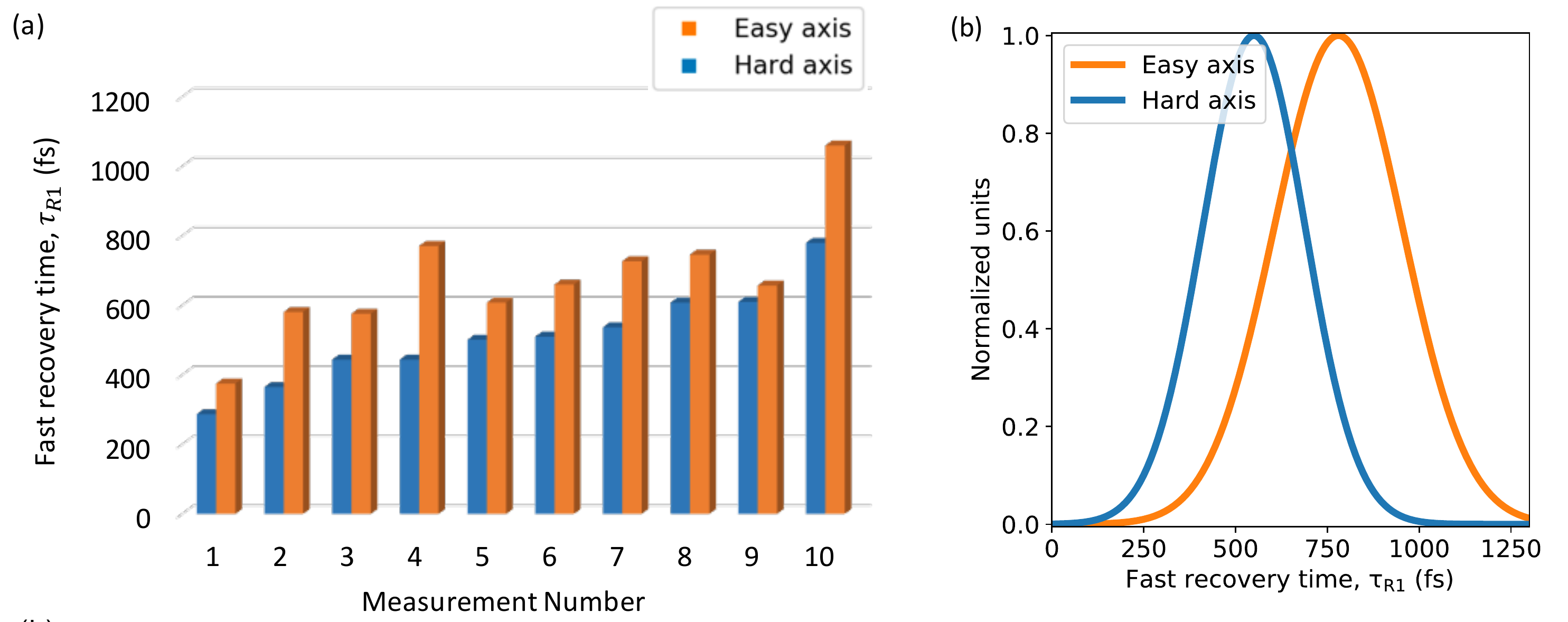}
\caption{\label{fig:Sup1} (a) Fast recovery time constant $\tau_{R1}$ of easy axis and hard axis for ten different measurements, with the absorbed fluence varying between 1 mJ/cm$^2$ to 2 mJ/cm$^2$. (b) Gaussian distributions of the fast recovery time constants for the measurements given in panel (a). The distribution for the easy axis is centered around 780 fs with a standard deviation of 175 fs whereas that for the hard axis is centered around 550 fs with a standard deviation of 140 fs. }
\end{figure*}

The robustness in the determination of $\tau_{R1}$ for easy and hard magnetization axes was checked with ten repeated measurements by varying the pump fluence in the range $1-2$ mJ/cm$^2$ for various regions of the samples the sample. These measurements are summarized in Fig.\ \ref{fig:Sup1}(a) in the form of an histogram plots and ordered by increasing $\tau_{R1}$ for the hard magnetization axis. Figure \ref{fig:Sup1}(b) shows a normalized Gaussian distribution of these recovery times for the easy and hard axes orientations, whose half width half maximum is the standard deviation of the ten repeated measurements. This plot is included to show schematically that even by completely neglecting the fluence dependence, the extracted time constants for the hard magnetization axis cluster around a substantially lower value than the time constants for the easy magnetization axis. This strongly prove the robustness of our results. We calculated the ratio of easy axis to hard axis recovery time to be 1.35, and the standard deviation of the ratio to be 0.19, as reported in the main text. We also performed measurements with four different combinations of pump and probe polarization. Irrespective of these combinations, we obtained the same trend in the results as the ones reported here.

\begin{table*}[!h]
\caption{\label{tab:table1}Table Extracted fit values using Eq.\ (\ref{eq:one}) for selected fluence values. }
\begin{ruledtabular}
\begin{tabular}{ccccccc}
 \multicolumn{2}{c}{}&\multicolumn{2}{c}{\hspace{0.75 cm}Amplitude (\%)}&\multicolumn{3}{c}{\hspace{0.3 cm}Relaxation time}\\
 Fluence (mJ/cm$^2$)&Easy axis / Hard axis&A$_1$&A$_2$&$\tau_{m}$ (fs)&$\tau_{R1}$ (fs)&$\tau_{R2}$ (ps)\\ \hline
 1& Easy &-22&-5.5&130&350&15 \\
  & Hard &-21&-5.5&130&260&10 \\
 1.5& Easy &-26&-7&130&520&17\\
  & Hard &-27&-8&130&360&12\\
 2& Easy &-33&-10&130&750&20\\
  & Hard &-34&-10&130&480&13\\
 4& Easy &-60&-18&180&730&23\\
  & Hard &-59&-17&130&540&18\\
\end{tabular}
\end{ruledtabular}
\end{table*}

\clearpage
\onecolumngrid 

\section{Structural and magnetic characterization of the sample}

In this section, we present the measurements used to characterize the crystalline structure, magneto-crystalline anisotropy and magneto-optical properties of the sample. The sample stack consists of 3 nm of platinum as the cap layer, 15 nm of hcp-cobalt and 5 nm of chromium as the seed layer with MgO as the substrate.

\begin{figure}[h!]
\includegraphics[width=0.8\columnwidth]{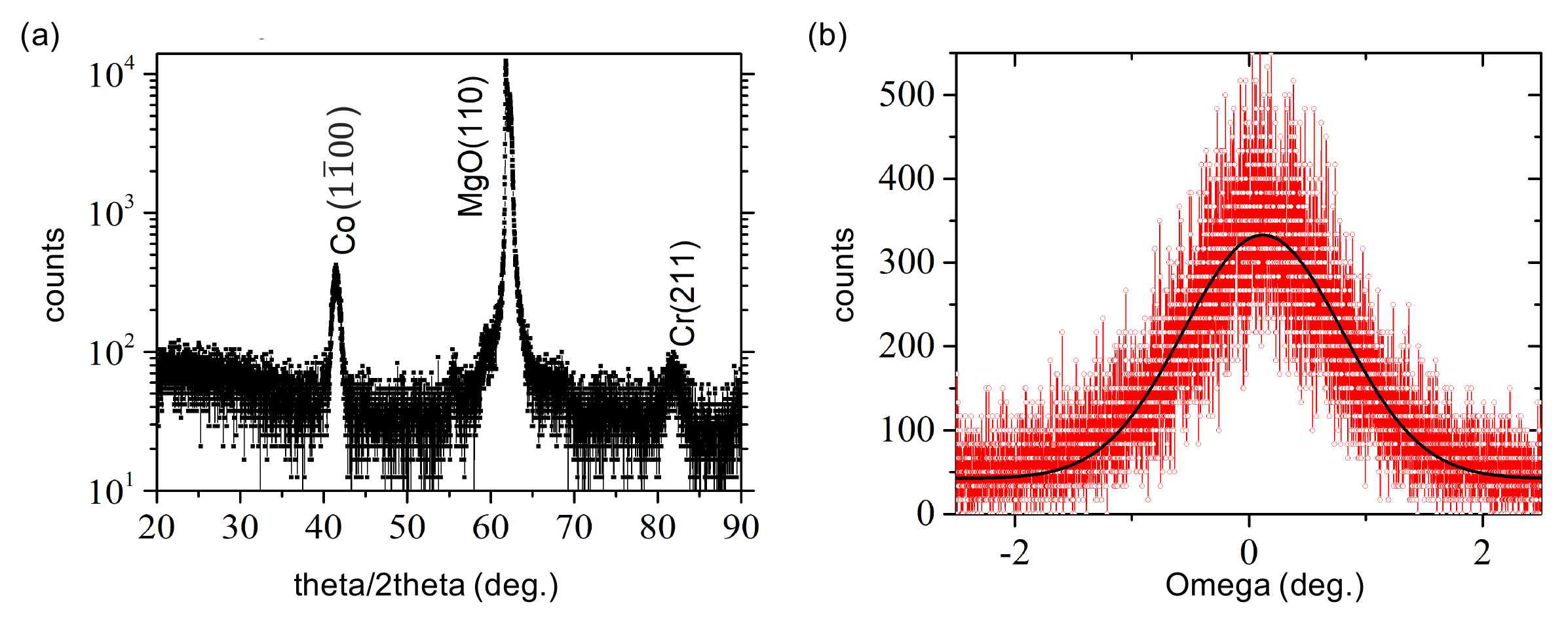}
\caption{\label{fig:Sup5}XRD characterization of the sample. (a) 2$\theta$ scan for out of plane characterization. (b) Rocking curve of cobalt layer for in plane characterization. The FWHM of the rocking curve is about 1.5 degree.}
\end{figure}

\begin{figure}[h!]
\includegraphics[width=0.8\columnwidth]{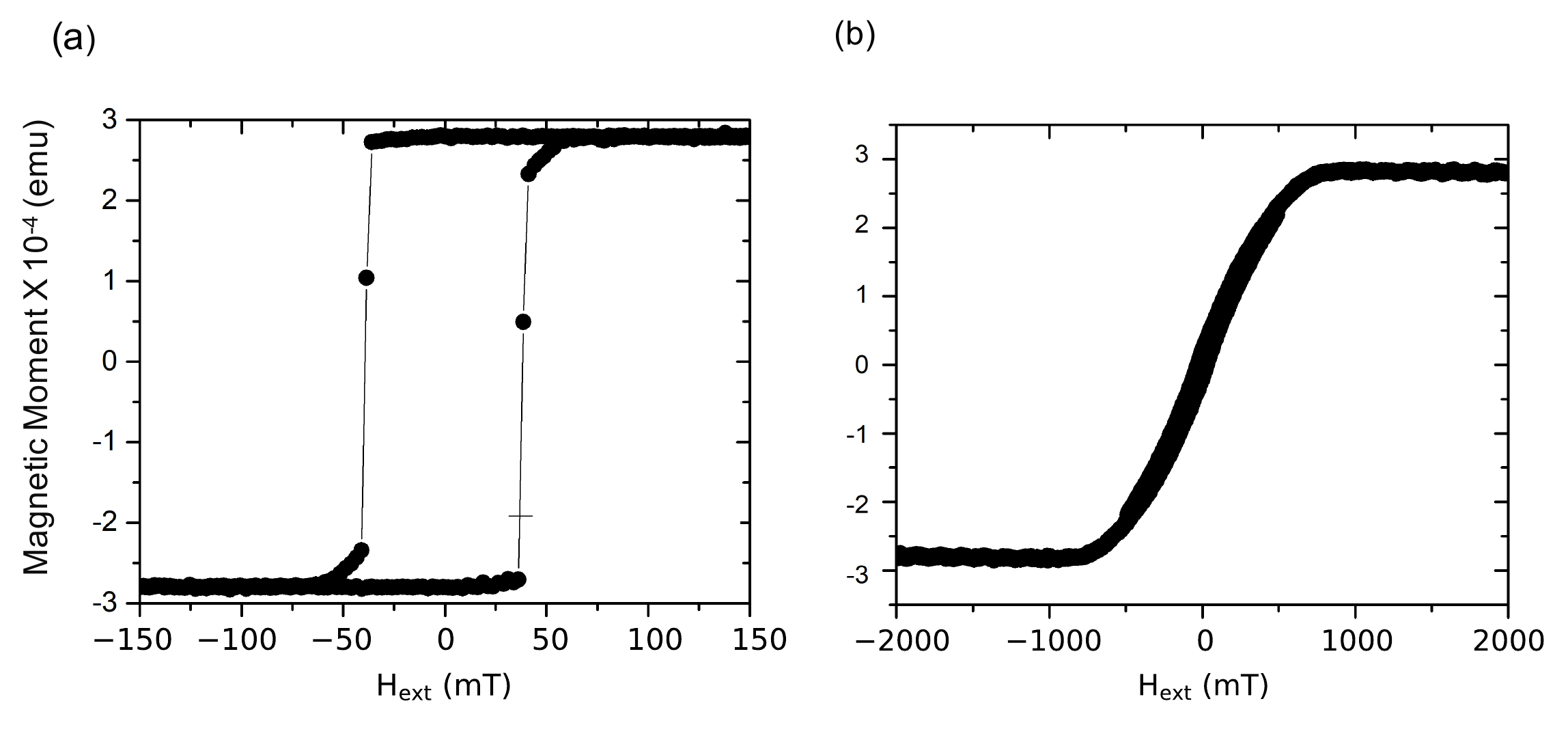}
\caption{\label{fig:Sup3} Magnetic characterization of the sample using a VSM. Magnetization loops measured in the plane of the sample along (a) the easy axis of magnetization [0001] and (b) the hard axis of magnetization [11$\bar{2}$0].}
\end{figure}

\begin{figure}[h!]
\includegraphics[width=0.8\columnwidth]{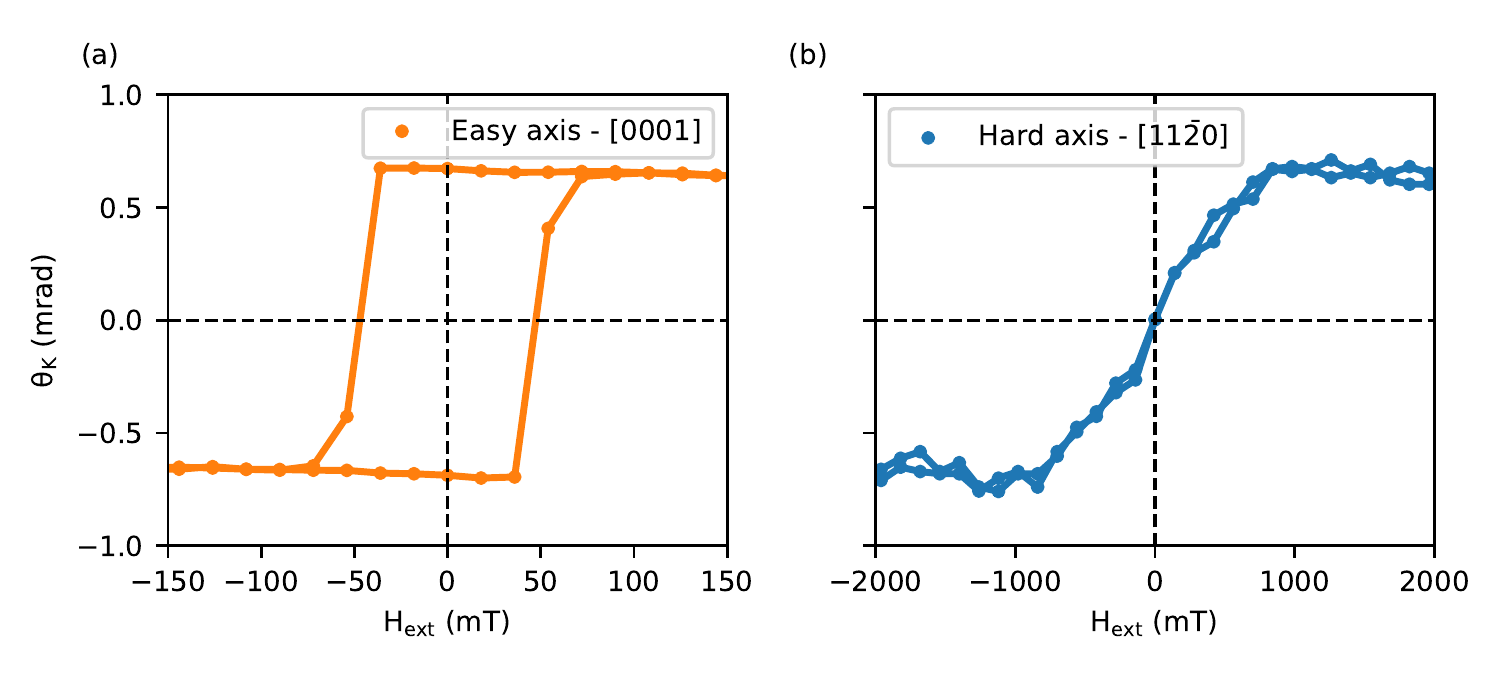}
\caption{\label{fig:Sup4} Characterization of the sample using the magneto-optical Kerr effect. Kerr rotations measured in the plane of the sample along (a) the easy axis of magnetization [0001] and (b) the hard axis of magnetization [11$\bar{2}$0].}
\end{figure}

Fig. \ref{fig:Sup5} shows the out-of-plane and the in-plane  structural characterization using X-Ray Diffraction (XRD). Panel (a) of the figure shows that the c-axis is in the sample plane, while panel (b) is the rocking curve. Figure \ref{fig:Sup3} shows the saturation magnetization for the easy and hard axes orientations, measured using a Vibrating Sample Magnetometer (VSM). A bias field of approximately $\pm$50 mT is required to saturate the sample along easy axis, whereas a field of $\pm$800 mT is required to saturate it along the hard axis direction. 
The MOKE magnetometry measurements for the easy and hard axes orientations are given in Fig.\ \ref{fig:Sup4}. We used \textit{p}-polarized, 800 nm low-intensity femtosecond optical pulses for this measurement. They have the same qualitative shape of the magnetization loops measured using the VSM. The Kerr rotation corresponding to the saturation magnetization is 0.7 mrad. This value was used as M$_0$ to calculate the relative change of magnetization in Eq.\ (\eqref{eq:one}).

\onecolumngrid

\clearpage
\section{Pump-probe measurements: additional data}
\begin{figure}[hbt!]
\includegraphics[width=0.7\columnwidth]{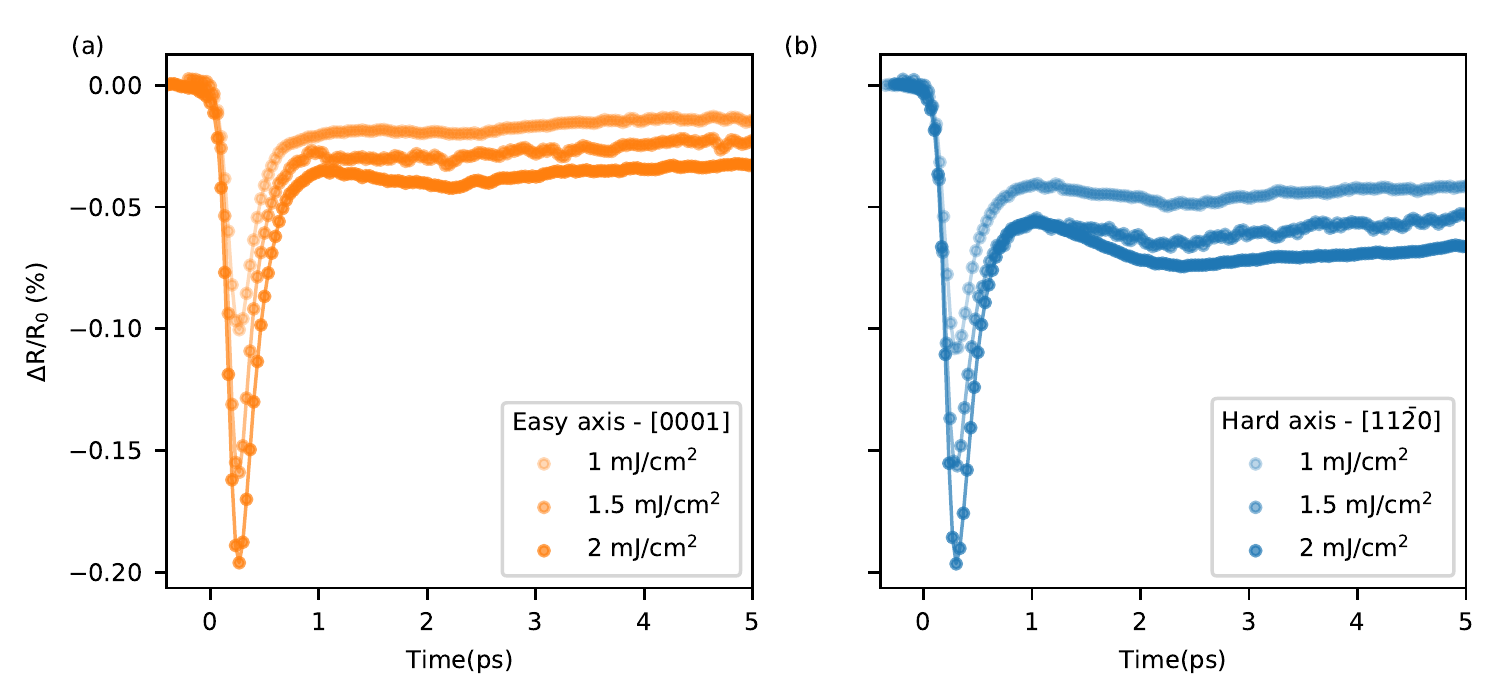}
\caption{\label{fig:Sup2} Transient reflectivity dynamics measured along the (a) easy [0001] and (b) hard [11$\bar{2}$0] magnetization axes. The pump is \textit{s}-polarized and the probe is \textit{p}-polarized. The calculated fluence absorbed by the film is given in the legend.}
\end{figure}

Fig. \ref{fig:Sup2} shows the pump-induced reflectivity change in the sample. We measured the change in reflectivity of \textit{p}-polarized probe following the pump excitation. The maximum observed change of reflectivity is less than 0.2\% when pumped with a pulse of fluence 2 mJ/cm$^2$, whereas the relative change in the magneto-optical response was two orders of magnitude larger.

\begin{figure}[h!]
\includegraphics[width=0.7\columnwidth]{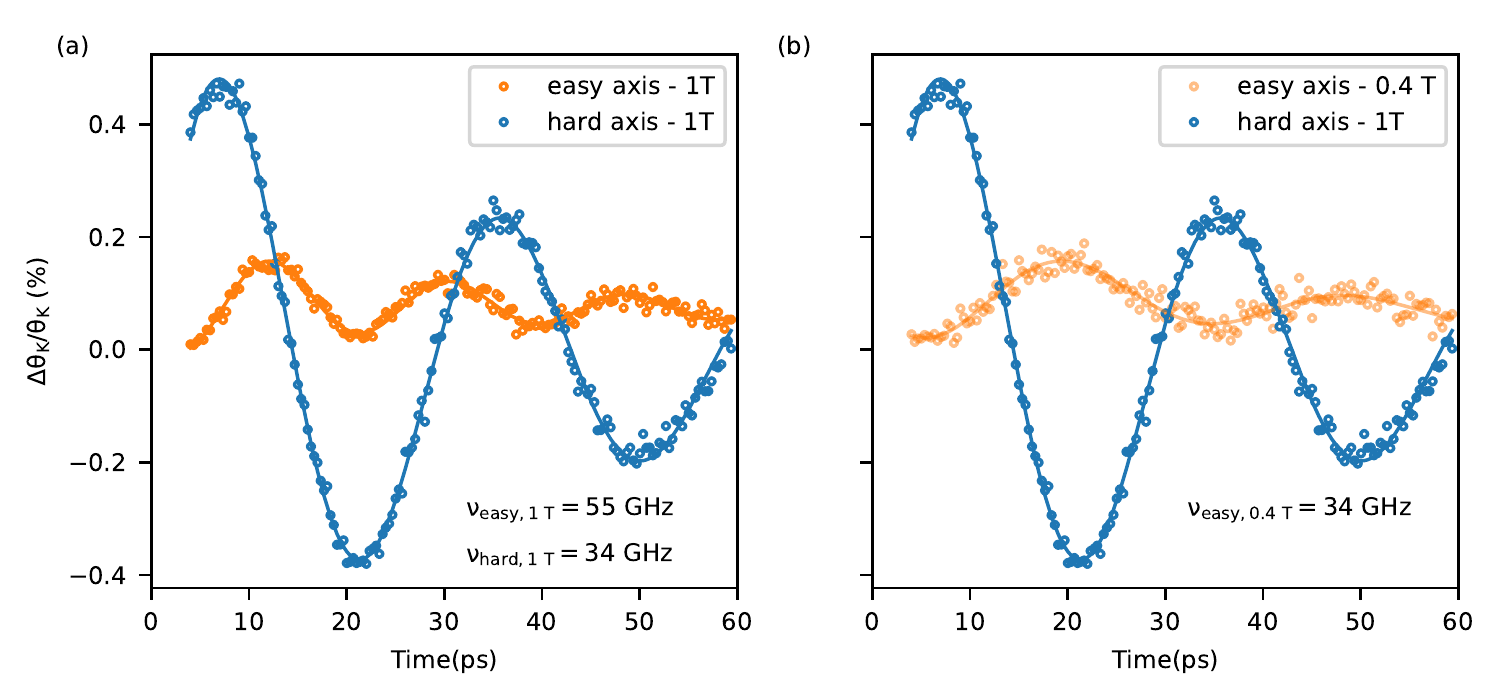}
\caption{\label{fig:Sup6}FMR measurements for easy and hard axes of magnetization when (a) external bias field of 1T for both directions, and (b) when the easy axis is saturated with 0.4 T and the hard axis with 1 T bias fields.}
\end{figure}

We also observed and characterized the FMR of the sample. Fig.\  \ref{fig:Sup6}(a) shows FMR oscillations for easy and hard axes orientations with a bias field of 1 T. The observed FMR frequency of the hard axis is 34 GHz and that of the easy axis is 55 GHz. This difference in FMR frequency between easy and hard axes is understood in terms of the different effective fields along the two directions, caused by the magneto-crystalline anisotropy of cobalt, which is estimated to be approximately 0.6 T \cite{cullity2011introduction}. In order to check the consistency of our arguments, we repeated the easy axis measurement with a lower saturation bias field of 0.4 T, which in addition to the anisotropy field would give 1 T, and thus it is expected to give approximately the same FMR frequency as for the hard axis saturated at 1 T. Figure \ref{fig:Sup6}(b) shows the comparison of these two measurements, and indeed the extracted FMR frequency is approximately the same (34 GHz) for both orientations. We also repeated the measurements with various bias fields, all above the saturation field for the respective axis, and observed that none of our results (in particular the extraction of the fast recovery time constant $\tau_{R1}$) are sensitive to the bias field value.

\clearpage
\section{Calculated phonon dispersions}

\begin{figure*}[!hbt]
\includegraphics[width=0.8\columnwidth]{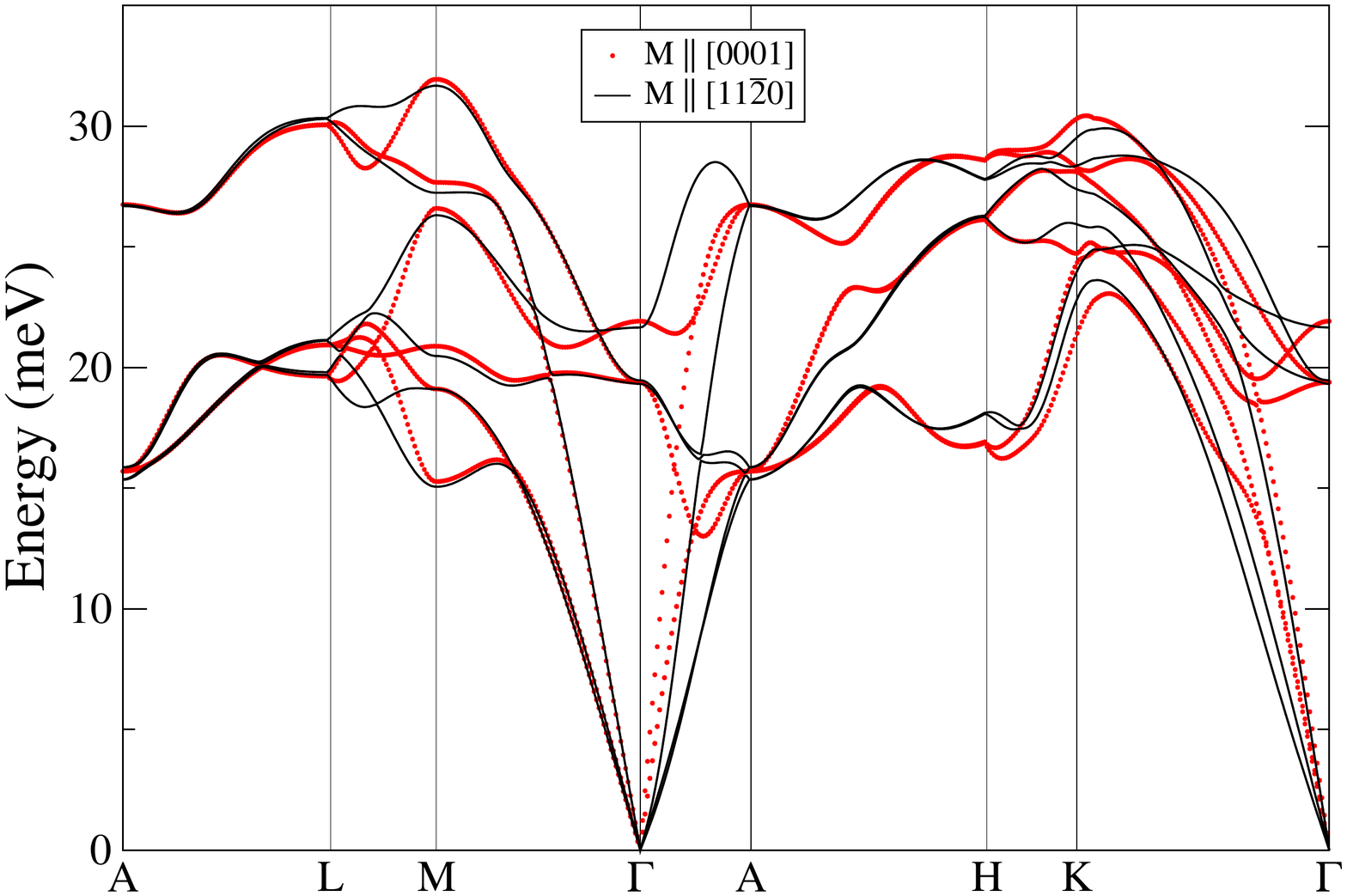}
\caption{\label{fig:spectra} \textit{Ab initio} calculated phonon dispersions of hcp Co, for magnetization $M$ either along the easy magnetization axis ([0001]) or along the hard axis ([11$\bar{2}$0]).}
\end{figure*}

The phonon dispersions of hcp Co were calculated \textit{ab initio}, using the first-principles formalism outlined in the main text. In Fig.\ \ref{fig:spectra}  we show the computed phonon dispersions along high-symmetry lines in the hcp Brillouin zone, for the magnetization either along the hard or the easy magnetization axis. 

\nocite{}
\bibliography{supplemental.bib}